\documentclass[12pt]{article}

\usepackage{graphicx}
\usepackage{amsfonts}

\def\gtwid{\mathrel{\raise.3ex\hbox{$>$\kern-.75em\lower1ex\hbox{$\sim$}}}}
\def\ltwid{\mathrel{\raise.3ex\hbox{$<$\kern-.75em\lower1ex\hbox{$\sim$}}}}
\def\square{\kern1pt\vbox{\hrule height 1.2pt\hbox{\vrule width 1.2pt\hskip 3pt
   \vbox{\vskip 6pt}\hskip 3pt\vrule width 0.6pt}\hrule height 0.6pt}\kern1pt}

\begin{document}

\begin{titlepage}

\begin{flushright}
CCTP-2014-09 \\
UFIFT-QG-14-04
\end{flushright}

\vskip 1cm

\begin{center}
{\bf A Caveat on Building Nonlocal Models of Cosmology}
\end{center}

\vskip 1cm

\begin{center}
N. C. Tsamis$^{1*}$ and R. P. Woodard$^{2\dagger}$
\end{center}

\vskip .5cm

\begin{center}
\it{$^{1}$ Institute of Theoretical Physics \& Computational Physics, \\
Department of Physics, University of Crete, \\
GR-710 03 Heraklion, HELLAS}
\end{center}

\begin{center}
\it{$^{2}$ Department of Physics, University of Florida,\\
Gainesville, FL 32611, UNITED STATES}
\end{center}

\vspace{.5cm}

\begin{center}
ABSTRACT
\end{center}

Nonlocal models of cosmology might derive from graviton loop corrections 
to the effective field equations from the epoch of primordial inflation.
Although the Schwinger-Keldysh formalism would automatically produce 
causal and conserved effective field equations, the models so far proposed 
have been purely phenomenological. Two techniques have been employed to
generate causal and conserved field equations: either varying an invariant
nonlocal effective action and then enforcing causality by the ad hoc 
replacement of any advanced Green's function with its retarded counterpart, 
or else introducing causal nonlocality into a general ansatz for the field 
equations and then enforcing conservation. We point out here that the two 
techniques access very different classes of models, and that neither one of 
them may represent what would actually arise from fundamental theory.

\begin{flushleft}
\end{flushleft}

\vskip 1cm

\begin{flushleft}
$^{*}$ e-mail: tsamis@physics.uoc.gr \\
$^{\dagger}$ e-mail: woodard@phys.ufl.edu
\end{flushleft}

\end{titlepage}

\section{Introduction}\label{intro}

Although quantum corrections to the effective field equations are known to
be nonlocal, this is not typically a macroscopic effect for two reasons:
\begin{enumerate}
\item{The virtual particles circulating in the loops are massive; and}
\item{When massless virtual particles do participate, they either lack 
self-interactions or else these self-interactions have a high dimension
which renders them ineffective on large scales.}
\end{enumerate}
The nonlocal effects of perturbative quantum field theory derive from 
inverse differential operators, and those of massive particles can be 
expanded in a series of local, higher derivatives,
\begin{equation}
\frac{i}{\partial^2 \!-\! m^2} = -\frac{i}{m^2} \Bigl[ 1 + 
\frac{\partial^2}{m^2} + \frac{\partial^4}{m^4} + \dots \Bigr] \; .
\end{equation}
The massless photons of quantum electrodynamics (QED) do produce observable
macroscopic effects \cite{Bloch:1937pw}, but these reduce to slight 
rescalings of ``hard'' (i.e., without corrections from low virtual momenta) 
rates and cross sections because photons lack self-interactions 
\cite{Weinberg:1965nx}. The same result is true for gravitons with zero
cosmological constant because their self-interactions are of dimension five
($h \partial h \partial h$) \cite{Weinberg:1965nx}. Long-range graviton 
corrections to the Newton \cite{Donoghue:1993eb,Donoghue:1994dn} and Coulomb 
\cite{BjerrumBohr:2002sx} potentials are suppressed for the same reason,
\begin{equation}
\Phi(r) = \Phi_{\rm classical} \Bigl\{ 1 + \frac{\# G}{r^2} + \dots \Bigr\} 
\; .
\end{equation}
($G$ is Newton's constant.) Macroscopic effects from (nearly) massless 
neutrinos are even smaller because the weak interaction has dimension six 
\cite{Feinberg:1968zz,Hsu:1992tg}.

Quantum chromodynamics (QCD) consists, in perturbation theory, of quarks and
massless gluons with self-interactions of dimension four. In this case infrared 
effects are so strong that the macroscopic spectrum of hadrons and baryons is 
entirely different from the quarks and gluons of perturbation theory. One could 
presumably follow the causal evolution of this transformation --- at least its 
opening stages --- by releasing a prepared state of free QCD vacuum and 
employing the Schwinger-Keldysh formalism \cite{Schwinger:1960qe,
Mahanthappa:1962ex,Bakshi:1962dv,Bakshi:1963bn,Keldysh:1964ud,Chou:1984es,
Jordan:1986ug,Calzetta:1986ey}. Precisely this sort of evolution must have 
occurred during the early universe around the time of the QCD phase transition.

When a nonzero cosmological constant is included, quantum gravity consists of
massless gravitons with a self-interaction of {\it dimension three}. It ought
therefore to suffer even stronger infrared effects than QCD. Veneziano showed
that the infrared effects of a massless scalar with a dimension three coupling
are so strong that the Bloch-Nordsieck procedure fails to produce finite 
rates and cross sections \cite{Veneziano:1972rs}. What happens instead is that
the vacuum decays, and one can indeed follow the first stages of this process 
using the Schwinger-Keldysh formalism \cite{Tsamis:1994ca}.
 
Polyakov early suggested that infrared effects in quantum gravity with a positive
cosmological constant $\Lambda$ should be so strong that they screen the bare 
$\Lambda$ \cite{Polyakov:1982ug}. Of course explicit computation of anything in
quantum gravity is very difficult, the more so in this case because one requires 
the ultraviolet finite part and because the appropriate background geometry is de 
Sitter rather than flat space. However, what computations have been made 
\cite{Tsamis:1996qm,Tsamis:1996qk,Tsamis:2005je,Miao:2005am,Miao:2006gj,
Miao:2007az,Kahya:2007bc,Kahya:2007cm,Miao:2008sp,Miao:2012bj,Leonard:2013xsa,
Mora:2013ypa,Glavan:2013jca} do include a number of effects \cite{Tsamis:1996qm,
Tsamis:1996qk,Miao:2005am,Miao:2006gj,Miao:2007az,Miao:2008sp,Miao:2012bj,
Leonard:2013xsa,Mora:2013ypa,Glavan:2013jca} that show secular corrections which 
grow like $G \Lambda \ln[a(t)]$, where $a(t)$ is the scale factor
\cite{Tsamis:2005hd,Prokopec:2007ak}. Over the course of a prolonged phase of de 
Sitter expansion the factor of $\ln[a(t)]$ must eventually overwhelm the small 
loop-counting parameter of $G \Lambda$, at which point perturbation theory breaks 
down. 

These considerations have prompted a proposal for simultaneously resolving the 
problem of the cosmological constant and providing a natural model of primordial 
inflation \cite{Tsamis:1996qq,Tsamis:2011ep}. The proposal is based on three 
contentions:
\begin{itemize}
\item{That the bare cosmological constant is not unreasonably small;}
\item{That this triggered primordial inflation; and}
\item{That inflation was brought to an end by the gradual accumulation of 
self-gravitation between the infrared virtual gravitons which are ripped out of 
the vacuum by inflation.}
\end{itemize}
The third item on this list is controversial \cite{Garriga:2007zk,Tsamis:2008zz},
however, there seems little doubt that inflation does create an ensemble of 
infrared gravitons \cite{Grishchuk:1974ny}. This is what caused the primordial 
tensor spectrum \cite{Starobinsky:1979ty} which the BICEP2 detector has recently 
claimed to resolve \cite{Ade:2014xna}. It is difficult to understand why these
gravitons would not attract one another, at least a little, or how the effect, 
which starts from zero, can avoid growing stronger as more of the newly created
gravitons come into causal contact with one another. Indeed, one can easily show 
that only the time lag imposed by causality prevents an inflating universe from 
experiencing gravitational collapse after fewer than ten e-foldings 
\cite{Tsamis:2011ep}!

As plausible as quantum gravitational back-reaction might seem, there is no 
simple way of analytically following the process by which it might stop
inflation, and hence no simple way of making testable predictions. Just
accessing the initial stages of the process requires a 2-loop computation, 
which represents a year's effort \cite{Tsamis:1996qm}, and will only show 
fractional corrections to the expansion rate of the form $- (G \Lambda)^2 
\ln[a(t)]$ \cite{Tsamis:2011ep}. The full series of leading infrared 
logarithms takes the form \cite{Tsamis:2005hd,Prokopec:2007ak},
\begin{equation}
H(t) = \sqrt{\frac13 \Lambda} \; \Bigg\{ 1 - G \Lambda \sum_{\ell = 2}^{\infty}
\Bigl( G \Lambda \ln\Bigl[\frac{a(t)}{a(t_i)}\Bigr]\Bigr)^{\ell-1} \Biggr\} 
\; , \label{series} 
\end{equation}
where $t_i$ is the beginning of inflation. From this series we see that 
perturbation theory breaks down when $\ln[a(t)/a(t_i)] \sim 1/G\Lambda$. 
Evolving beyond this point would require a nonperturbative resummation 
technique. There seems to be some chance that such a technique can be devised
because Starobinsky and Yokoyama were able to find one for scalar potential 
models which also exhibit infrared logarithms 
\cite{Starobinsky:1994bd,Woodard:2005cw,Tsamis:2005hd}. Their technique has 
been generalized to scalars which interact with photons \cite{Prokopec:2007ak}, 
and to scalars which interact with fermions \cite{Miao:2006pn}, but the 
generalization to quantum gravity has not yet been accomplished 
\cite{Miao:2008sp,Kitamoto:2010et,Kitamoto:2011yx}. 

In the absence of a nonperturbative resummation technique one approach has been
to explore simple ans\"atze for the most cosmologically significant part of the 
effective field equations \cite{Tsamis:1997rk}. These effective field equations 
must be nonlocal in order to recover known secular dependence of 
the perturbative result (\ref{series}) because the de Sitter limit of any local 
curvature degenerates to factors of $\Lambda$ times sums of products of the 
metric. In addition to attempting to represent the quantum gravitational 
back-reaction to primordial inflation \cite{Tsamis:2009ja,Tsamis:2010ph,
Romania:2012av}, nonlocal modifications of gravity have also been invoked to 
describe the current phase of cosmic acceleration without recourse to dark 
energy \cite{Deser:2007jk,Deffayet:2009ca,Park:2012cp,Deser:2013uya,
Ferreira:2013tqn,Dodelson:2013sma,Woodard:2014iga}, to
provide a metric-based realization \cite{Soussa:2003vv,Hehl:2008eu,Hehl:2009es,
Blome:2010xn,Deffayet:2011sk,Arraut:2013qra,Deffayet:2014lba,Woodard:2014wia} 
of Milgrom's MOdified Newtonian Dynamics (MOND) \cite{Milgrom:1983ca,
Milgrom:1983pn,Milgrom:1983zz}, and to solve a variety of other problems 
\cite{nonloc}.

One problem with nonlocal modifications of gravity is generating field 
equations which are both causal and conserved. If the modification was derived 
from fundamental theory these two requirements would follow naturally from the
Schwinger-Keldysh formalism. However, because no such derivation is currently
possible, two entirely phenomenological approaches have been followed instead:
\begin{enumerate}
\item{Proceeding from a general causal ansatz for the field equations and
then determining free functions to enforce conservation \cite{Tsamis:2009ja,
Tsamis:2010ph,Romania:2012av}; or}
\item{Varying an invariant action whose nonlocality derives from inverse 
differential operators --- which ensures conservation --- and then enforcing
causality by replacing all advanced Green's functions with their retarded
cousins \cite{Soussa:2003vv,Deser:2007jk,Deser:2013uya,Woodard:2014iga,
Woodard:2014wia}.}
\end{enumerate}
The point of this work is to demonstrate that the two approaches do not 
access the same range of models, and that neither may include the sort of 
models which would actually come from the Schwinger-Keldysh formalism.

This paper has five sections of which the first is coming to an end. In 
section \ref{models} we present general models of the two types, assuming
their nonlocality derives from the inverse scalar d'Alembertian acting on 
the Ricci scalar. In section \ref{conflict} we demonstrate that no choice 
of the various free functions will make the models agree for more than a
single expansion history. Section \ref{realthing} contrasts the nonlocality
of these simple models with what one actually gets from perturbative 
corrections in the Schwinger-Keldysh formalism. Our conclusions comprise
section \ref{epilogue}.

\section{Two Classes of Models}\label{models}

The point of this section is to present general representatives from the
two classes of models described above in section \ref{intro}. We make the
additional requirement that their nonlocality derives from acting on the 
Ricci scalar with the inverse of the scalar covariant d'Alembertian, 
\begin{equation}
\square \equiv \frac1{\sqrt{-g}} \partial_{\mu} \Bigl( \sqrt{-g} g^{\mu\nu}
\partial_{\nu} \Bigr) \; . \label{square}
\end{equation}
We define the inverse of $\square$ with retarded boundary conditions. In
each case we give the model's correction $\Delta G_{\mu\nu}$ to the 
classical Einstein tensor, so that the effective field equations of pure
gravity take the form,
\begin{equation}
G_{\mu\nu} + \Delta G_{\mu\nu}[g] = -\Lambda g_{\mu\nu} \; . \label{GReqns} 
\end{equation}
We also specialize each class of models to the homogeneous, isotropic and 
spatially flat geometry relevant to cosmology,
\begin{equation}
ds^2 = -dt^2 + a^2(t) d\vec{x} \!\cdot\! d\vec{x} \quad \Longrightarrow 
\quad H(t) \equiv \frac{\dot{a}}{a} \; . \label{FRW}
\end{equation}

\subsection{Perfect Fluid-Based Models}\label{fluid}

The first class of models is based on assuming that the most cosmologically 
relevant part of the quantum gravitational stress tensor takes the perfect
fluid form \cite{Tsamis:2009ja,Tsamis:2010ph,Romania:2012av}, 
\begin{equation}
T_{\mu\nu}[g] = \Bigl( \rho[g] + p[g]\Bigr) u_{\mu}[g] u_{\nu}[g] + p[g] 
g_{\mu\nu} \quad , \quad g^{\mu\nu} u_{\mu}[g] u_{\nu}[g] = -1 \; . 
\label{Tmn}
\end{equation}
Of course this makes the correction to the Einstein tensor,
\begin{equation}
\Delta G_{\mu\nu}[g] = -8\pi G T_{\mu\nu}[g] \; .
\end{equation}
We take the pressure to be a general function of $\frac1{\square} R$,
consistent with the series (\ref{series}) of perturbative leading 
logarithms,
\begin{equation}
p[g] = \Lambda^2 f\Bigl( -G \Lambda \frac1{\square} R\Bigr) \; . 
\label{pressure}
\end{equation}
The energy density $\rho[g]$ and 4-velocity $u_{\mu}[g]$ are then 
determined by conservation. It can be shown \cite{Tsamis:2009ja} that all 
models of this type experience a long period of inflation, followed by a 
phase of oscillations, provided only that the function $f(Z)$ grows
monotonically and without bound. If matter couplings are added, which
permit the dissipation of energy, it becomes plausible that the 
oscillatory epoch ends in a normal epoch of radiation domination. At 
this point $R = 0$, and $\frac1{\square} R$ becomes constant, so 
$8\pi G T_{\mu\nu} = +\Lambda g_{\mu\nu}$ \cite{Tsamis:2010ph}.

In the cosmological geometry (\ref{FRW}) the only nonzero components of
the affine connection are,
\begin{equation}
\Gamma^i_{~j0} = H \delta^{i}_{j} \qquad , \qquad \Gamma^0_{ij} = 
H g_{ij} \; . \label{Gammas}
\end{equation}
The nonzero components of the Riemann tensor are,
\begin{equation}
R^0_{~i0j} = (\dot{H} + H^2) g_{ij} \qquad , \qquad R^i_{~jk\ell} =
H^2 \Bigl( \delta^i_k g_{j\ell} - \delta^i_{\ell} g_{jk} \Bigr) \; .
\label{Riemmann}
\end{equation}
Tracing produces the nonzero components of the Ricci tensor, and tracing
again gives the Ricci scalar,
\begin{equation}
R_{00} = -3 (\dot{H} + H^2) \quad , \quad R_{ij} = (\dot{H} + 3 H^2) 
g_{ij} \quad , \quad R = 6 \dot{H} + 12 H^2 \; . \label{Ricci}
\end{equation}
It follows that the two nonzero components of the Einstein tensor are,
\begin{equation}
G_{00} = 3 H^2 \qquad , \qquad G_{ij} = -(2 \dot{H} + 3 H^2) g_{ij} 
\; . \label{Einstein}
\end{equation}
When acted on an arbitrary function of time $F(t)$, the inverse scalar
d'Alem\-ber\-tian takes the form,
\begin{equation}
\frac1{\square} F = -\int_{t_i}^t \!\! \frac{dt'}{a^3(t')} \!\! 
\int_{t_i}^{t'} \!\! dt'' a^3(t'') F(t'') \; . \label{1/Box}
\end{equation}

In the cosmological geometry (\ref{FRW}) the pressure is,
\begin{equation}
p(t) = \Lambda^2 f\Biggl( G \Lambda \!\! \int_{t_i}^t \!\! 
\frac{dt'}{a^3(t')} \!\! \int_{t_i}^{t'} \!\! dt'' a^3(t'')
\Bigl[ 6 \dot{H}(t'') \!+\! 12 H^2(t'')\Bigr] \Biggr) \; .
\label{press}
\end{equation}
Enforcing conservation implies the 4-velocity field and energy
density are, 
\begin{equation}
u_{\mu} = -\delta^0_{\mu} \quad , \quad 
\rho(t) = - p(t) + \frac1{a^3(t)} \!\! \int_{t_i}^t \!\! dt' a^3(t') 
\dot{p}(t') \; . \label{urho} 
\end{equation}
Hence the nonzero components of $\Delta G_{\mu\nu}$ are,
\begin{equation}
\Delta G_{00} = -8 \pi G \rho \qquad , \qquad \Delta G_{ij} = - 8\pi G
p g_{ij} \; . \label{model1}
\end{equation}

\subsection{Action-Based Models}\label{action}

The second class of models consists of a Lagrangian which is a general 
function of $\frac1{\square} R$,
\begin{equation}
\Delta \mathcal{L} = \Lambda^2 h\Bigl( -G \Lambda \frac1{\square} R\Bigr) 
\sqrt{-g} \; . \label{DL}
\end{equation}
Varying the integral of (\ref{DL}) produces conserved but acausal field 
equations. Causality is enforced, without disturbing conservation, by 
the ad hoc replacement of every advanced Green's function with its 
retarded cousin \cite{Soussa:2003vv,Deser:2007jk,Deser:2013uya,
Woodard:2014iga,Woodard:2014wia},
\begin{equation}
\Delta G_{\mu\nu}[g] \equiv \frac{16 \pi G}{\sqrt{-g}} \Biggl( 
\frac{\delta \Delta S[g]}{\delta g^{\mu\nu}} \Biggr)_{{\rm avd} 
\rightarrow {\rm ret}} \; . \label{IBPtrick}
\end{equation}
This works because conservation depends only on the relation $\square 
\cdot \frac1{\square} = 1$, which is obeyed by both advanced and 
retarded Green's functions.

A quicker way to derive $\Delta G_{\mu\nu}[g]$ is to follow Nojiri and
Odintsov \cite{Nojiri:2007uq} in localizing the theory through the 
introduction of auxiliary scalar fields. Our model (\ref{DL}) requires
two scalars: $\phi$ to stand for $\frac1{\square} R$, and a Lagrange
multiplier $\xi$ whose variation implies $\square \phi = R$. The 
localized Lagrangian density is,
\begin{equation}
\Delta \mathcal{L} \longrightarrow \Lambda^2 h\Bigl(-G \Lambda \phi\Bigr)
\sqrt{-g} - \Bigl[ \partial_{\mu} \xi \partial_{\nu} \phi g^{\mu\nu} + 
\xi R\Bigr] \sqrt{-g} \; . \label{locDL}
\end{equation}
The scalar field equations are,
\begin{eqnarray}
\frac1{\sqrt{-g}} \frac{\delta \Delta S}{\delta \xi} & = & \square \phi
- R = 0 \; , \label{phieqn} \\
\frac1{\sqrt{-g}} \frac{\delta \Delta S}{\delta \phi} & = & \square \xi
-G \Lambda^3 h'\Bigl(-G \Lambda \phi\Bigr) = 0 \; . \label{xieqn}
\end{eqnarray}
The correction to the Einstein tensor is,
\begin{eqnarray}
\lefteqn{ \Delta G_{\mu\nu} = 8\pi G \Biggl\{ \Bigl[-\Lambda^2 h\Bigl(-
G \Lambda \phi\Bigr) + \partial_{\rho} \xi \partial_{\sigma} \phi 
g^{\rho\sigma} \Bigr] g_{\mu\nu} } \nonumber \\
& & \hspace{4cm} - 2 \partial_{(\mu} \xi \partial_{\nu)} \phi - 2
\Bigl[ G_{\mu\nu} + g_{\mu\nu} \square - D_{\mu} D_{\nu} \Bigr] \xi
\Biggr\} \; , \qquad \label{DGmn}
\end{eqnarray}
where $D_{\mu}$ stands for the covariant derivative and parenthesized 
indices are symmetrized. 

It is clear from (\ref{locDL}) that one linear combination of the 
auxiliary scalars would be a ghost if they were truly independent 
dynamical variables \cite{Deser:2013uya,Woodard:2014iga}. To avoid this
we follow the same procedure that was successfully invoked to avoid ghosts
in another nonlocal cosmology model which is based on $\frac1{\square} R$
\cite{Deser:2013uya,Woodard:2014iga}. The trick is just to define each 
scalar with vanishing initial value data so that we can express the
solutions to equations (\ref{phieqn}-\ref{xieqn}) in terms of the retarded
Green's function,
\begin{equation}
\phi[g] \equiv \frac1{\square} R \qquad , \qquad \xi[g] \equiv 
\frac{G \Lambda^3}{\square} h'\Bigl(-G \Lambda \phi[g]\Bigr) \; .
\label{phixi}
\end{equation}
Substituting (\ref{phixi}) in (\ref{DGmn}) gives the same causal,
conserved and ghost-free effective field equations that follow from the 
partial integration trick (\ref{IBPtrick}).

It remains to specialize relations (\ref{DGmn}) and (\ref{phixi}) to the
geometry (\ref{FRW}) of cosmology. The auxiliary scalars are,
\begin{eqnarray}
\phi(t) & = & -\int_{t_i}^{t} \!\! \frac{dt'}{a^3(t')} \!\! \int_{t_i}^{t'} 
\!\! dt'' a^3(t'') \Bigl[ \dot{H}(t'') \!+\! 12 H^2(t'')\Bigr] \; , 
\label{phiFRW} \\
\xi(t) & = & -G \Lambda^3 \!\! \int_{t_i}^{t} \!\! \frac{dt'}{a^3(t')} 
\!\! \int_{t_i}^{t'} \!\! dt'' a^3(t'') h'\Bigl( -G\Lambda \phi(t'')\Bigr)
\; . \qquad \label{xiFRW}
\end{eqnarray}
The nontrivial components of $\Delta G_{\mu\nu}$ are,
\begin{eqnarray}
\Delta G_{00} & = & 8\pi G \Bigl[ \Lambda^2 h - \dot{\xi} \dot{\phi} 
- 6 H \dot{\xi} - 6 H^2 \xi \Bigr] \; , \label{model2a} \\
\Delta G_{ij} & = & -8\pi G \Bigl[ \Lambda^2 h + 2 G \Lambda^3 h' +
\dot{\phi} \dot{\xi} + 2 H \dot{\xi} - (4 \dot{H} \!+\! 6 H^2) \xi\Bigr]
g_{ij} \; . \qquad \label{model2b}
\end{eqnarray}

\section{Why and How They Disagree}\label{conflict}

Each of the models described in section \ref{models} depends upon an arbitrary
function of the nonlocal quantity $Z \equiv -G\Lambda \frac1{\square} R$. If 
these models represent the same physics then it must be possible to define the
function $f(Z)$ of the perfect fluid model in terms of the function $h(Z)$ of
its action-based analog so that both models give the same $\Delta G_{\mu\nu}$.
We would need this agreement to hold not only for a general cosmological 
geoemtry (\ref{FRW}) but also for perturbations around this geometry. However, 
it isn't even possible to make the two classes of models agree for a general 
expansion history $a(t)$. To see this, note that getting the same pressure from
expressions (\ref{model1}) and (\ref{model2b}) requires,
\begin{eqnarray}
\lefteqn{ f\Bigl( -G \Lambda \phi(t) \Bigr) = h\Bigl(-g\Lambda \phi(t)\Bigr)
+ 2 G \Lambda h'\Bigl(-G \Lambda \phi(t) \Bigr) } \nonumber \\
& & \hspace{2.3cm} + \frac1{\Lambda^2} \Biggl\{ \dot{\phi}(t) \dot{\xi}(t) + 2
H(t) \dot{\xi}(t) - \Bigl[4 \dot{H}(t) \!+\! 6 H^2(t) \Bigr] \xi(t) \Biggr\}
\; . \qquad \label{ffromh}
\end{eqnarray}
The terms on the first line are no problem but those on the second line 
preclude general agreement between the two models because they depend upon
the Hubble parameter and its derivatives, as well as on derivatives and 
integrals of what should be the single independent variable $Z(t) = -G
\Lambda \phi(t)$. We can choose the relation between $f(Z)$ and $h(Z)$ to make 
the models coincide for one particular expansion history $a(t)$ but they will 
not agree for other expansion histories.

To see the problem in more detail, let us work out the relation between $f(Z)$ 
and $h(Z)$ which is needed to make the models agree for the de Sitter expansion 
history $a(t) = e^{H_i t}$, where $3 H_i^2 = \Lambda$. Evaluating expression 
(\ref{phiFRW}) for de Sitter reveals that it is an excellent approximation,
after a long period of inflation, to regard $\phi(t)$ as a linear function of 
time,
\begin{equation}
\phi(t) = -4 H_i (t \! -\! t_i) + \frac43 - \frac43 e^{-3 H_i (t - t_i)} 
\approx - 4 H_i \Delta t \; . \label{phidS}
\end{equation} 
Substituting this approximation into expression (\ref{xiFRW}) and assuming 
that the $t''$ integration is dominated by the growth of $a^3(t'')$ results in 
an equally valid approximation for $\xi(t)$,
\begin{equation}
\xi(t) \approx -\frac14 \Lambda h\Bigl(-G\Lambda \phi(t)\Bigr) \; . \label{xidS}
 \end{equation}
Similar approximations for the derivatives are,
\begin{equation}
\dot{\phi}(t) \approx -4 H_i \qquad , \qquad \dot{\xi}(t) \approx -G \Lambda^2
H_i h'\Bigl(-4 G \Lambda \phi(t)\Bigr) \; . \label{dStimeds}
\end{equation} 
Substituting relations (\ref{phidS}, (\ref{xidS}) and (\ref{dStimeds}) in
(\ref{ffromh}) implies that the two models will agree for de Sitter provided,
\begin{equation}
f(Z) \approx \frac32 h(Z) + \frac83 G \Lambda h'(Z) \; . \label{dSagree}
\end{equation}

Suppose relation (\ref{dSagree}) pertains, which will make the two models 
agree during de Sitter inflation. Now consider an expansion history with a 
long period of nearly de Sitter inflation, followed by an intermediate 
phase in which the Ricci scalar becomes negative, and then a long period of
perfect radiation domination with $R(t) = 0$,
\begin{eqnarray}
t_i < t < t_1 & \Longrightarrow & {\rm Inflation\ with} \; R(t) > 0 \; , \\
t_1 < t < t_2 & \Longrightarrow & {\rm Oscillation\ with} \; R(t) < 0 \; , \\
t_2 < t < t_3 & \Longrightarrow & {\rm Radiation\ with} \; R(t) = 0 \; .
\end{eqnarray}
This is an important expansion history because the perfect fluid model (with
matter) follows it for any function $f(Z)$ which increases monotonically
and without bound \cite{Tsamis:2009ja}. During the phase of radiation 
domination we can write,
\begin{equation}
t_2 < t < t_3 \Longrightarrow \phi(t) = -\!\!\int_{t_i}^{t_2} \!\!\!\!\! 
\frac{dt'}{a^3(t')} \!\! \int_{t_i}^{t'} \!\!\!\! dt'' a^3(t'') R(t'') - 
\!\! \int_{t_2}^{t} \!\!\!\! \frac{dt'}{a^3(t')} \!\! \int_{t_i}^{t_2} 
\!\!\!\! dt'' a^3(t'') R(t'') . \label{phirad}
\end{equation}
We can obviously choose the expansion history $a(t)$ to make the rightmost
integral of (\ref{phirad}) vanish because this represents only a single 
condition on the infinite number of points at which $a(t'')$ can be 
specified for $t_i < t'' < t_2$. Suppose this has been done, which makes 
$\phi(t)$ constant and its first derivative zero,
\begin{equation}
t_2 < t < t_3 \Longrightarrow \phi(t) = -\!\!\int_{t_i}^{t_2} \!\! 
\frac{dt'}{a^3(t')} \!\! \int_{t_i}^{t'} \!\!\!\! dt'' a^3(t'') R(t'') 
\equiv \phi_{\rm cr} \quad , \quad \dot{\phi}(t) = 0 \; . \label{phispec}
\end{equation}
This means that the perfect fluid pressure becomes exactly constant
during radiation domination,
\begin{equation}
t_2 < t < t_3 \Longrightarrow p(t) = \Lambda^2 f\Bigl( -G\Lambda 
\phi_{\rm cr} \Bigr) \; . \label{pres1}
\end{equation}

It is easy to see that the pressure of the action-based model during the
epoch of radiation domination differs from (\ref{pres1}). Note first that, 
if $H(t_2) \equiv H_2$ and $a(t_2) \equiv a_2$, then we have,
\begin{equation}
t_2 < t < t_3 \Longrightarrow H(t) = \frac{H_2}{1 \!+\! 2 H_2 (t \!-\! 
t_2)} \qquad , \qquad a(t) = a_2 \Bigl[1 \!+\! 2 H_2 (t \!-\! t_2)
\Bigr]^{\frac12} \; . \label{Haspec}
\end{equation}
Substituting expressions (\ref{phispec}) and (\ref{Haspec}) into 
(\ref{xiFRW}) gives the following late time form for $\xi(t)$,
\begin{equation}
t_2 < t < t_3 \Longrightarrow \xi(t) = -G \Lambda^3 h'\Bigl(-G \Lambda
\phi_{\rm cr} \Bigr) \times \frac{t^2}{5} + O(t) \; . \label{latexi}
\end{equation}
We infer from relation (\ref{ffromh}) that agreement between the two 
models for $t_2 < t < t_3$ requires,
\begin{equation}
f\Bigl( -G\Lambda \phi_{\rm cr}\Bigr) = h\Bigl(-G \Lambda \phi_{\rm cr}
\Bigr) + \frac32 G \Lambda h'\Bigl(-G \Lambda \phi_{\rm cr}\Bigr) \; .
\label{fhrad}
\end{equation}
Subtracting (\ref{fhrad}) from (\ref{dSagree}) gives,
\begin{equation}
\frac12 h\Bigl(-G\Lambda \phi_{\rm cr}\Bigr) + \frac76 G \Lambda
h'\Bigl(-G\Lambda \phi_{\rm cr}\Bigr) = 0 \; ,
\end{equation}
which is not obeyed for a general function $h(Z)$. Many, many similar
disagreements can be derived.

\section{Schwinger-Keldysh Field Equations}\label{realthing}

No one knows what form the Schwinger-Keldysh effective field equations 
of quantum gravity might take beyond perturbation theory and for an 
arbitrary cosmological geometry (\ref{FRW}). However, quite a bit of 
experience has been gained working at one and two loop orders on de 
Sitter background \cite{Tsamis:1996qq,Tsamis:2005je,Miao:2006gj,
Kahya:2007cm,Miao:2008sp,Mora:2013ypa,Glavan:2013jca,Park:2011kg,
Leonard:2014zua}; that is, with $a = e^{H_i t}$ in the cosmological
background (\ref{FRW}). The point of this section is to discuss three 
potentially significant differences between the nonlocality manifest 
in these equations and the $1/\square$ nonlocality so far explored in 
phenomenological models:
\begin{enumerate}
\item{The effective field equations typically involve nonlinear powers
of inverse differential operators;}
\item{One consequence of this nonlinearity is that the effective field
equations typically involve the real part of the propagator as well as
the retarded Green's function; and}
\item{The effective field equations typically involve inverse
tensor differential operators in addition to the inverse scalar
d'Alembertian.}
\end{enumerate}

\begin{figure}
\vskip -.5cm
\includegraphics[width=4.5cm,height=2.25cm]{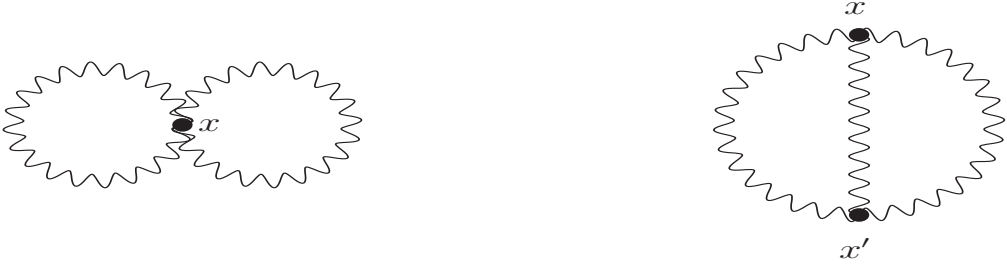}
\vskip 1.2cm
\caption{Two of the many two loop diagrams which contribute
to the 1PI graviton 1-point function \cite{Tsamis:1996qm}. Wavy
lines represent graviton propagators.}
\label{grav1pt}
\end{figure}

To simplify the discussion of the first two issues we will suppress 
the indices of the graviton field, its propagator, and its interaction 
vertices: $h_{\mu\nu}(x) \rightarrow h(x)$. In this notation the 
effective field equations take the form,
\begin{equation}
\sum_{n=1}^{\infty} \frac1{(n-1)!} \!\! \int \!\! d^4x_2 h(x_2) 
\ldots \int \!\! d^4x_n h(x_n) \Gamma^{(n)}(x,x_2,\dots,x_n) = 0 \; ,
\label{effeqns}
\end{equation}
where $\Gamma^{(n)}(x_1,x_2,\dots,x_n)$ is the full 
one-particle-irreducible (1PI) n-point function, including the 
classical contributions. Figure~\ref{grav1pt} shows two of the many 
diagrams which contribute to the 1PI graviton 1-point function at
two loop order \cite{Tsamis:1996qm}. The diagram on the left involves
the product of two coincident propagators,
\begin{equation}
\Gamma^{(1a)}(x) = \frac{\kappa^4}{2} \Bigl[ i\Delta(x;x)\Bigr]^2
\; , \label{fig1a}
\end{equation}
where $\kappa^2 \equiv 16\pi G$ is the loop-counting parameter of
quantum gravity and $i\Delta(x;x')$ is the propagator. The diagram on
the right of Fig.~\ref{grav1pt} involves an integral of the product 
of three propagators,
\begin{equation}
\Gamma^{(1b)}(x) = -\frac{i \kappa^4}{3!} \!\! \int \!\! d^Dx'
\Bigl[ i\Delta(x;x') \Bigr]^3 \; . \label{fig1b}
\end{equation}
Of course propagators are the inverses of differential operators so
expressions (\ref{fig1a}-\ref{fig1b}) manifest the nonlinearity 
which was point 1 above. Both diagrams are divergent so they must
be evaluated in $D$ spacetime dimensions, before being combined with
the appropriate counterterms to give the $D \rightarrow 4$ limit
which appears in the full 1-point function.

\begin{figure}
\vskip -.5cm
\includegraphics[width=4.0cm,height=2.0cm]{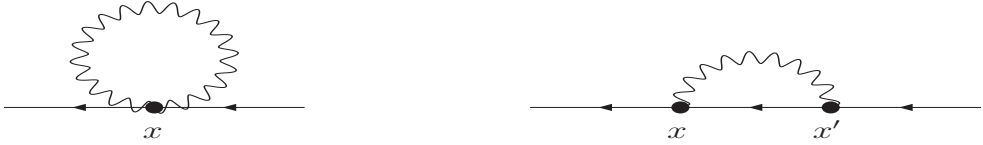}
\caption{The two primitive one loop diagrams which contribute
to the 1PI fermion 2-point function \cite{Miao:2005am}. Wavy lines
represent graviton propagators while solid lines with arrows stand
for the fermion propagator.} 
\label{ferm2pt}
\end{figure}

In reality the vertices of quantum gravity contain derivatives and
index structures which preclude more than a single one of the 
propagators in expressions (\ref{fig1a}-\ref{fig1b}) from contributing 
an infrared logarithm. The part of the graviton propagator which 
potentially contributes an infrared logarithm is the same as the 
propagator of a massless, minimally coupled scalar
\cite{Onemli:2002hr,Onemli:2004mb},
\begin{eqnarray}
\lefteqn{i\Delta(x;x') = \frac{H^{D-2}}{(4\pi)^{\frac{D}2}} \Biggl\{
\frac{\Gamma(\frac{D}2)}{\frac{D}2 \!-\! 1} \Bigl( \frac4{y}
\Bigr)^{\frac{D}2 -1} \!\!\!\!\!+ \frac{\Gamma(\frac{D}2 \!+\! 1)}{
\frac{D}2 \!-\! 2} \Bigl(\frac4{y}\Bigr)^{\frac{D}2 -2} \!\!\!\!\! + 
\frac{\Gamma(D \!-\! 1)}{\Gamma(\frac{D}2)} \ln(a a') + K} \nonumber \\
& & \hspace{.5cm} + \sum_{n=1}^{\infty} \Biggr[ \frac1{n}
\frac{\Gamma(n \!+\! D \!-\! 1)}{\Gamma(n \!+\! \frac{D}2)}
\Bigl(\frac{y}4 \Bigr)^n - \frac1{n \!-\! \frac{D}2 \!+\! 2}
\frac{\Gamma(n \!+\! \frac{D}2 \!+\! 1)}{\Gamma(n \!+\! 2)}
\Bigl(\frac{y}4 \Bigr)^{n - \frac{D}2 +2} \!\Biggr] \Biggr\} ,
\qquad \label{prop}
\end{eqnarray}
where $K$ is a $D$-dependent constant and $y$ is the de Sitter
length function,
\begin{equation}
y(x;x') \equiv a a' \Bigl[H_i^2  \Vert \vec{x} \!-\! \vec{x}' \Vert^2
- (\vert e^{-H_i t} \!-\! e^{-H_i t'} \vert \!-\! i \epsilon)^2 
\Bigr] \; .
\end{equation}
A detailed study \cite{Miao:2008sp} of graviton corrections to the 
propagation of massless fermions (see Fig.~\ref{ferm2pt}) concluded 
that infrared logarithms derive entirely from the 2nd and 3rd terms 
on the first line of expression (\ref{prop}) ,
\begin{equation}
\frac{\Gamma(\frac{D}2 \!+\! 1)}{
\frac{D}2 \!-\! 2} \Bigl(\frac4{y}\Bigr)^{\frac{D}2 -2} \!\!\!\!\! + 
\frac{\Gamma(D \!-\! 1)}{\Gamma(\frac{D}2)} \ln(a a') \; .
\label{IRlogs}
\end{equation}
The second term of (\ref{IRlogs}) is responsible for the famous 
secular dependence of the coincidence limit \cite{Vilenkin:1982wt,
Linde:1982uu,Starobinsky:1982ee} which occurs in the left hand diagrams 
of Figures \ref{grav1pt} and \ref{ferm2pt}. The first term of 
(\ref{IRlogs}) vanishes in the dimensionally regulated coincidence 
limit, but it can and does contribute an infrared logarithm in the 
diagram to the right of Fig. \ref{ferm2pt}. One of the reasons it has 
been so difficult to sum the series of leading infrared logarithms is 
that multiplicative constants of order one derive from {\it all} parts 
of the various propagators which do not contribute infrared logarithms, 
so they cannot be simplified.

One might worry that the infrared logarithms from coincident propagators
such as those of expression (\ref{fig1a}) do not seem to be associated 
with $\frac1{\square}$ acting on anything. However, Dolgov and Pelliccia 
have shown --- and for a general metric, not just de Sitter --- that 
$i\Delta(x;x)$ is proportional to $\frac1{\square}$ acting on the trace
of the free scalar stress tensor \cite{Dolgov:2005se}. Of course the
relation between the graviton propagator and the scalar propagator is
only valid in the cosmological geometry (\ref{FRW}) \cite{Lifshitz:1945du}
but this is precisely the class of geometries of interest. It is not known
what the trace of the scalar stress tensor should be for a general 
cosmological geometry. This quantity has the same dimension as the product 
of two curvatures, and the models of section \ref{models} were based on 
assuming it is $-\Lambda \times R$. A later class of models 
\cite{Tsamis:2010ph} explored the possibility that it is $R_{00} \times 
R$. We shall argue elsewhere that the most interesting part of the trace 
cannot be the product of any local curvatures \cite{Romania:2014XX}.

As noted previously, the phenomenological models so far explored employ 
ad hoc techniques to derive casual and conserved field equations. This 
is not at all what fundamental theory predicts. The effective field
equations of fundamental theory derive from a variant of the usual
Feynman rules which is known as the Schwinger-Keldysh formalism 
\cite{Schwinger:1960qe,Mahanthappa:1962ex,Bakshi:1962dv,Bakshi:1963bn,
Keldysh:1964ud,Chou:1984es,Jordan:1986ug,Calzetta:1986ey}. We shall omit
a general explanation of this technique and merely describe how it 
affects expression (\ref{fig1b}). Schwinger-Keldysh propagators have 
$\pm$ polarities at each of their two endpoints. The $++$ polarity 
corresponds to the usual Feynman propagator whose real and imaginary 
parts we can write as,
\begin{equation}
i\Delta_{\scriptscriptstyle ++}(x;x') = R(x;x') + i I(x;x') \; .
\end{equation}
Recall that the real part of the propagator depends upon the quantum 
vacuum state and is generally nonzero throughout spacetime, whereas 
the imaginary part is the sum of the advanced and retarded Green's
functions and is zero for spacelike separation. The other polarity we
require is $+-$,\footnote{The $--$ and $-+$ polarities are just the 
complex conjugates of the $++$ and $+-$ polarities, respectively.}
\begin{equation}
i\Delta_{\scriptscriptstyle +-}(x;x') = R(x;x') 
- i {\rm sgn}(t \!-\! t') I(x;x') \; .
\end{equation}
Note that the $++$ and $+-$ polarities agree for $t < t'$, whereas they
are complex conjugates for $t > t'$.

With this explanation we can now discuss point 2 above. It turns out 
that the Schwinger-Keldysh result for expression (\ref{fig1b}) is
\cite{Tsamis:1996qm},
\begin{eqnarray}
\Gamma^{(1b)}_{\rm SK}(x) & = & -\frac{i \kappa^4}{3!} \!\! \int \!\!
d^Dx' \Biggl\{ \Bigl[i\Delta_{\scriptscriptstyle ++}(x;x') \Bigr]^3 -
\Bigl[i\Delta_{\scriptscriptstyle +-}(x;x') \Bigr]^3 \Biggr\} \; , 
\qquad \label{int1} \\
& = & \frac{\kappa^4}{3!} \!\! \int \!\! d^Dx' \theta(t \!-\! t')
\Biggl\{ 6 R^2(x;x') I(x;x') - 2 I^3(x;x') \Biggr\} \; . \label{int2}
\end{eqnarray}
Only points ${x'}^{\mu}$ on or within the past lightcone of $x^{\mu}$ make
nonzero contributions to expression (\ref{int2}) because each term of the 
integrand involves at least one factor of the imaginary part of the 
Feynman propagator, which vanishes for spacelike separation. When 
multiplied by $\theta(t - t')$ this imaginary part becomes the retarded
Green's function. The phenomenological models so far studied all derive
their nonlocality from the retarded Green's function, but it is evident
from expression (\ref{int2}) that {\it the real part of the propagator
can also contribute.} This might be important because the real part of 
the propagator depends more strongly on infrared gravitons than does the 
imaginary part.

We come finally to point 3. Ferreira and Maroto have made a 
pioneering study of nonlocality from inverting higher spin d'Alembertians 
\cite{Ferreira:2013tqn}. They noted that these inverses generically 
involve exponentially growing modes, however, this can be easily fixed by 
adding the same nonminimal couplings which nature provides for the quanta 
in question. In Lorentz gauge, the photon kinetic operator is not the 
vector d'Alembertian but rather \cite{Allen:1985wd,Tsamis:2006gj},
\begin{equation}
(D_1)_{\mu}^{~\nu} \equiv \square_{\mu}^{~\nu} - R_{\mu}^{~\nu} \; .
\label{vector}
\end{equation}
Under the assumptions of homogeneity and isotropy (\ref{FRW}) vector 
fields must take the form, $V_{\mu} = V_0(t) \delta^0_{\mu}$. Hence the
inverse $F_{\mu} \equiv (\frac1{D_1} V)_{\mu}$ obeys,
\begin{equation}
V_0(t) = -\ddot{F}_0 - 3 H \dot{F}_0 - 3 \dot{H} F_0 = -\frac{d}{dt}
\Bigl[ \frac1{a^3} \frac{d}{dt} \Bigl( a^3 F_0\Bigr) \Bigr] \; .
\label{vectoreqn}
\end{equation}
Equation (\ref{vectoreqn}) is straightforward to solve with retarded 
boundary conditions,
\begin{equation}
\Bigl( \frac1{D_1} V\Bigr)_{\mu} = -
\frac{\delta^0_{\mu}}{a^3(t)} \!\! \int_{t_i}^t \!\! dt' a^3(t') \!\!
\int_{t_i}^{t'} \!\! dt'' V_0(t'') \; . \label{vectorinv}
\end{equation}

The same sort of fix works for higher spins. In de Donder gauge the 
graviton kinetic operator is not the tensor d'Alembertian but rather 
\cite{Miao:2011fc},
\begin{equation}
(D_2)_{\mu\nu}^{~~ \rho\sigma} \equiv \square_{\mu\nu}^{~~ \rho\sigma}
- 2 R_{\mu ~~ \nu}^{~ (\rho ~ \sigma)} + 2 R_{(\mu}^{~~ (\rho} 
\delta_{\nu)}^{~~ \sigma)} - \frac43 R \delta_{\mu}^{~ (\rho} 
\delta_{\nu}^{~ \sigma)} + \frac13 R g_{\mu\nu} g^{\rho\sigma} \; .
\label{tensor}
\end{equation}
The same cosmological symmetries of homogeneity and isotropy restrict 
any tensor to the form,
\begin{equation}
T_{00} = T_{00}(t) \quad , \quad T_{0i} = 0 = T_{i0} \quad , \quad
T_{ij} = T(t) g_{ij} \; .
\end{equation}
Hence $F_{\mu\nu} \equiv (\frac1{D_2} T)_{\mu\nu}$ obeys,
\begin{equation}
(D_2)_{\mu\nu}^{~~ \rho\sigma} F_{\rho\sigma} = T_{\mu\nu} \qquad 
\Longrightarrow \qquad \cases{ -\frac1{a^3} \frac{d}{dt} [ a^3
\dot{F}_{00} ] = T_{00} \; , \cr
-\frac1{a^3} \frac{d}{dt} [ a^3 \dot{F} ] = T \; . }
\label{tensoreqn}
\end{equation} 
Of course this is just the scalar d'Alembertian acting on each component
\cite{Lifshitz:1945du} and so the retarded inverse can be expressed as,
\begin{equation}
\Bigl( \frac1{D_2} T\Bigr)_{\mu\nu} = -\!\! \int_{t_i}^t \!\! 
\frac{dt'}{a^3(t')} \!\! \int_{t_i}^{t'} \!\! dt'' a^3(t'') 
T_{\mu\nu}(t'') \; . \label{tensorinv}
\end{equation}
One might be tempted to conclude that there is no need to distinguish
between $(D_2)_{\mu\nu}^{~~ \rho\sigma}$ and the scalar d'Alembertian,
and this is probably correct for perfect fluid models in which the
effective field equations are posited directly. However, the distinction 
does matter for action-based models in which the field equations are
obtained by variation because the two differential operators do not 
agree for a general metric.

\section{Epilogue}\label{epilogue}

As explained in section \ref{intro}, our aim in nonlocal model-building 
is to guess the most cosmologically significant part of the effective 
field equations which might pertain after the self-gravitation between 
inflationary gravitons has become nonperturbatively strong. We know the
perturbative limit (\ref{series}), and we suspect the correct model must 
be simple, so finding a reasonable ansatz for it seems possible. Nor is 
this exercise doomed to forever remain speculative. The prediction of a 
compelling cosmology would be a strong indication that the correct model 
had been found, and knowing its form might well facilitate a derivation 
from fundamental theory, just as the stochastic formalism of Starobinsky 
and Yokoyama \cite{Starobinsky:1994bd} for scalar potential models was 
later derived \cite{Tsamis:2005hd,Woodard:2005cw}.

It is neither possible, nor even particularly desirable, to incorporate
all details of the actual effective field equations. However, it is 
crucial that our nonlocal ans\"atze be sufficiently general to recover 
their most cosmologically significant portions. In this regard it is 
disturbing that that the two techniques so far employed to generate 
models seem to access different cosmologies. Subsection \ref{fluid} 
discussed a perfect fluid ansatz based on an arbitrary function 
$f(-G\Lambda \frac1{\square} R)$, whereas subsection \ref{action} 
described a general class of actions based on a different arbitrary 
function $h(-G \Lambda \frac1{\square} R)$. In section \ref{conflict} 
we showed that one cannot choose $f(Z)$ in terms of $h(Z)$ so as to 
make the two models agree for more than a single expansion history. 
One way to understand the difference between the two classes of models 
is that the localized form (\ref{locDL}) of the action-based model 
includes a conformal coupling $\xi R\sqrt{-g}$ whose variation induces 
terms in the effective field equations which cannot be present in the 
perfect fluid model.

It is also disturbing that there seem to be significant differences 
between the existing phenomenological models and the known one and two 
loop contributions to the effective field equations on de Sitter 
background. In section \ref{realthing} we considered three properties 
of the Schwinger-Keldysh effective field equations which are not shared 
by the models:
\begin{itemize}
\item{They typically involve nonlinear powers of inverse differential
operators;}
\item{They typically involve the real parts of propagators, in 
addition to the imaginary parts which appear in the retarded Green's
function; and}
\item{They typically involve the inverses of tensor differential
operators.}
\end{itemize}
Some of these differences may be more apparent than substantive. For
example, although expressions (\ref{fig1a}) and (\ref{fig1b}) involve
two and three propagators, it is known that only one propagator in 
each case can provide the crucial secular dependence which is 
represented by $\frac1{\square}$ in the models. The other propagators 
need to be present, but they might be well approximated, for the 
purposes of cosmology, by simple functions of the curvature. And the 
close relation (\ref{tensorinv}) between $\frac1{\square}$ and its
tensor cousin $\frac1{D_2}$ in the cosmological geometry (\ref{FRW})
means that the distinction between the two operators only shows up in 
the form of some extra terms in the variational equations.

The overall conclusion for now is that model-builders should be aware 
of the potential for problems with certain ans\"atze. We suspect that
the action-based models are more likely to be correct than the perfect
fluid ones because the effective field equations of fundamental theory 
derive from varying the Schwinger-Keldysh effective action, which is 
closely related to the in-out effective action. One feature we can
already see that existing models fail to correctly describe is the
curvature-squared source upon which $\frac1{\square}$ acts. The 
existing models are based upon taking this source to be either 
$-\Lambda \times R$ \cite{Tsamis:2009ja} or $R_{00} \times R$ 
\cite{Tsamis:2010ph}, which agree for de Sitter but not in general.
Preliminary analysis shows that the actual source should not even be
local \cite{Romania:2014XX}.

\vskip 1cm

\centerline{\bf Acknowledgements}

We are grateful for conversation and correspondence with A. Barvinsky, 
C. Deffayet, S. Deser, A. Dolgov, G. Esposito-Farese, P. Ferreira and 
T. A. Jacobson. This work was partially supported by the 
European Union (European Social Fund, ESF) and Hellenic national funds 
through the Operational Program ``Education and Lifelong Learning" 
of the National Strategic Reference Framework (NSRF) under the 
``$\Theta\alpha\lambda\acute{\eta}\varsigma$'' action MIS-375734, 
under the ``$A\rho\iota\sigma\tau\epsilon\acute{\iota}\alpha$'' 
action, under the ``Funding of proposals that have received a 
positive evaluation in the 3rd and 4th Call of ERC Grant Schemes''; 
by NSF grant PHY-1205591, and by the Institute for Fundamental Theory 
at the University of Florida.

\end{document}